\documentclass[useAMS,usenatbib]{mn2e}
\usepackage{graphicx}

\newcommand{\ms}{~M$_\mathrm{\odot}$}

\newcommand{\be}{\begin{equation}}
\newcommand{\ee}{\end{equation}}
\title[Solving the LMC Carbon Star Mystery]{Third Dredge-up in Low Mass Stars: Solving the LMC Carbon Star Mystery}
\author[R.J. Stancliffe, R.G. Izzard and C.A. Tout]{Richard J. Stancliffe\thanks{E-mail:
rs@ast.cam.ac.uk}$^1$, Robert G. Izzard$^2$ and Christopher A. Tout$^1$\\
$^1$Institute of Astronomy, The Observatories, Madingley Road, Cambridge CB3 0HA \\ $^2$Carolune Institute for Quality
Astronomy, http://freya.phys.uniroma1.it/$\sim$rgi/ciqua/}
\begin{document}

\date{Accepted 0000 December 00. Received 0000 December 00; in original form 0000 October 00}

\pagerange{\pageref{firstpage}--\pageref{lastpage}} \pubyear{0000}

\maketitle

\label{firstpage}

\begin{abstract}
A long standing problem with asymptotic giant branch (AGB) star models has been their inability to produce the low-luminosity carbon stars in the Large and Small Magellanic Clouds. Dredge-up must begin earlier and extend deeper. We find this for the first time in our models of LMC metallicity. Such features are not found in our models of SMC metallicity.

The fully implicit and simultaneous stellar evolution code {\sc{stars}} has been used to calculate the evolution of AGB stars with metallicities of $\mathrm{Z}=0.008$ and $\mathrm{Z}=0.004$, corresponding to the observed metallicities of the Large and Small Magellanic Clouds, respecitively. Third dredge-up occurs in stars of 1\ms\ and above and carbon stars were found for models between 1\ms\ and 3\ms. We use the detailed models as input physics for a population synthesis code and generate carbon star luminosity functions. We now find that we are able to reproduce the carbon star luminosity function of the LMC without any manipulation of our models. The SMC carbon star luminosity function still cannot be produced from our detailed models unless the minimum core mass for third dredge-up is reduced by 0.06\ms.
\end{abstract}

\begin{keywords}
stars: evolution, stars: AGB, carbon stars
\end{keywords}

\section{Introduction}
Carbon stars are stars which show the features of carbon rich molecules such as C$_{2}$ and CN in their spectra indicating they have a surface carbon-to-oxygen ratio (by number) greater than unity. They are typically identified in photometric surveys using the colours from narrow band filters centred near $7800$\AA\ and $8100$\AA\ (Cook \& Aaronson 1989). The first filter focuses on a TiO absorption feature whilst the latter centres on a CN absorption feature. A census of carbon stars has been performed in the Large Magellanic Cloud (LMC) and Small Magellanic Cloud (SMC). In the LMC a total of 7750 carbon stars have been found; in the SMC 1707 are known (Groenewegen 2004). The distances to the LMC and SMC are well determined and so we are able to construct luminosity functions for the carbon stars in both these locations. We are thus presented with an opportunity to test our models of stellar evolution.

It is believed that two populations of carbon stars exist which reflect two different formation routes. These populations are referred to as {\it intrinsic} and {\it extrinsic}. The intrinsic carbon stars are believed to be thermally pulsing asymptotic giant branch (TP-AGB) stars. It has long been known that the unstable double-shell burning in AGB stars leads to the phenomenon of third dredge-up (TDUP). The ashes of helium burning are mixed into the convective envelope enhancing the surface abundance of carbon (Iben \& Renzini 1983). In low-mass stars repeated occurence of TDUP gives rise to carbon stars because the products of helium burning are carbon plus a little oxygen. However, for stars of around 4\ms\ or greater (the limit varying with metallicity), the temperature at the base of the convective envelope can become high enough for the CN cycle to operate. Carbon is converted into nitrogen via proton captures (e.g. Iben \& Renzini 1983). This prevents the conversion to a carbon star. If sufficient mass is stripped from the envelope of such a star then the hot-bottom burning can cease and a carbon star can then form. The extrinsic carbon stars are believed to be formed by the accretion of carbon-rich material from a more evolved donor (Van Eck et al. 1998).

There is a long-standing problem with forming intrinsic carbon stars. Detailed stellar evolution models have so far proved unable to produce dredge-up at low enough core masses and hence at low enough luminosities. When Iben formulated the carbon star mystery (Iben 1981) calculations showed that dredge-up did not occur for core masses below about 0.6\ms\ (e.g. Sackmann 1980, Wood \& Zarro 1981). This is also borne out by more recent calculations but leads to problems in reproducing the observed luminosity functions of the LMC and the SMC. Using the results of Karakas et al. (2002), Izzard et al. (2004) determined that the core masses at dredge-up would have to be lower by 0.07\ms.

Most modern stellar evolution codes do not solve the equations of stellar structure, nuclear burning and mixing simultaneously. Mixing is typically solved for after the structure and burning have been determined. Examples include the codes used by Straniero et al. (1997) and Herwig (2000). Karakas et al. (2002) employ a partial simultaneous approach. A structure iteration is followed by a mixing iteration before another structure iteration is made. In this way a model is converged. Initial work on a fully implicit and simultaneous solution of these equations (whereby the structure, burning and mixing are all treated together in the {\it same} iteration) together with an improved treatment of diffusive mixing, on the TP-AGB was done by Pols \& Tout (2001). More recently Stancliffe, Tout \& Pols (2004) found that it was possible to obtain dredge-up at lower core masses than was previously thought.

Following on from that work, the evolution of stars of between 1 and $6\,$\ms\ at $\mathrm{Z}=0.008$ and $\mathrm{Z}=0.004$ is investigated. The outcome of this detailed modelling is used as the input physics to a population synthesis code in order to compare the luminosity functions predicted by the data with the observed LMC and SMC carbon star luminosity functions (CSLF). In section 2 we discuss the stellar evolution code used and the details of the models. The results of the these calculations are presented in section 3. The population synthesis code is briefly discussed in section 4 where we present our luminosity functions.

\section{Evolution Calculations}
The stellar evolution code {\sc stars} developed by Eggleton (1971) and much updated (e.g. Pols et al. 1995) is a fully implicit code that solves the equations of stellar structure, nuclear burning and diffusive mixing {\it simultaneously}. This is notably different to the way most codes work with mixing usually treated separately. We use the version of the code developed by Stancliffe, Tout and Pols (2004) with its modifications for use on the TP-AGB. It has an improved treatment of mixing and a viscous mesh to aid numerical stability.

It is well known that the phenomenon of TDUP is sensitive to the numerical treatment of the convective boundary (Frost \& Lattanzio 1996). It must be ensured that the convective boundary remains stable (see Mowlavi 1999 for a detailed discussion). This is achieved in the current code by the use of an arithmetic mean for the diffusion coffecient (see Pols \& Tout 2001 for full details). The adaptive mesh of the code provides a mass resolution of around $2\times10^{-6}$\ms\ at the boundary of the convective envelope. This is comparable with the resolution used by Herwig (2000). An increase in the resolution does not change the amount of material dredged-up. The code does not treat the changes in energy associated with mixing material of different composition (Wood 1981). However Frost (1997) finds that if this physics is not included the efficiency of third dredge-up is reduced.

Models of 1, 1.5, 2, 3, 4, 5 and 6 M$_\odot$\ were evolved from the pre-main sequence up to and along the TP-AGB. No convective overshooting or mass loss was considered at any stage of the evolution. We use the formalism of B\"{o}hm-Vitense (1958) for convection with $\alpha=1.925$, based on calibration to a solar model. Models were evolved with 199 mesh points up to core helium burning. At this point the resolution was increased to 499 mesh points in order to facilitate the transition to 999 meshpoints just before the onset of the TP-AGB.

\subsection{The Core Helium Flash}
Stars of masses up to around 2.3\ms\ are expected to undergo a core helium flash. However, this phase of evolution is numerically demanding and the {\sc stars} code is currently not suitable for a calculation of the evolution through it. Instead a model of the desired mass is run from the pre-main sequence up to the helium flash. The hydrogen exhausted core mass and the envelope composition are recorded. A 3 M$_\odot$ model is then evolved from the pre-main sequence up to the point where helium ignites in the core. During this evolution helium burning reactions are allowed to produce energy but not consume helium. Once helium has ignited under non-degenerate conditions mass is stripped from the envelope and the core is allowed to grow until the model has the desired envelope mass and core mass. The envelope composition is then set to that of the pre-flash model. We are therefore assuming that the helium flash proceeds so rapidly that the core mass doesn't change and that there is no change in the envelope composition. These are both standard assumptions.

\section{Results}
We are able to evolve through a significant number of thermal pulses in all our models and third dredge-up is found to occur in them all, including the 1\ms\ models. At both metallicities hot-bottom burning is found to occur in models of 4\ms\ or greater and these stars would not become carbon stars without mass loss. If these stars do become carbon stars once hot-bottom burning has ceased they have much larger core masses and so contribute to the more luminous end of the CSLF. Eventually each evolution sequence terminates due to unresolved numerical issues but the models are sufficiently evolved for our purposes where we are mostly concerned with the early pulses of the lowest-mass stars.

\subsection{LMC models}
For the LMC models we form carbon stars in all models of 3\ms\ and below. Details of the models are given in Table~\ref{tab:LMCmods}. Note that the formation of carbon stars in these models is extremely rapid. It requires only two or three thermal pulses with TDUP. This is because of the low abundance of oxygen in these models: only a small amount of carbon needs to be dredged up for it to become more abundant than oxygen. The efficiency of dredge-up is defined through the paramter $\lambda = {\Delta M_\mathrm{DUP}}/{\Delta M_\mathrm{c}}$, where $\Delta M_\mathrm{DUP}$ is the mass of material dredged up and $\Delta M_\mathrm{c}$ is the amount by which the H-exhausted core grows in the preceeding interpulse. The maximum $\lambda$ value of the 1\ms\ model quoted in Table~\ref{tab:LMCmods} is unlikely to be a true representation of the maximum efficiency reached because numerical instabilities prevent us evolving this model to a steady state.

\begin{table}
\begin{center}
\begin{tabular}{c|cccccc}
Mass & M$_\mathrm{1TP}$ & M$_\mathrm{TDUP}$ & $\lambda_\mathrm{max}$ & N$_\mathrm{tot}$ & N$_\mathrm{TDUP}$ & N$_\mathrm{c}$ \\ \hline
1   & 0.54803 & 0.57262 & 0.169 &  8 & 6 & 2 \\
1.5 & 0.56610 & 0.57127 & 0.752 & 40 & 4 & 2 \\
2   & 0.56419 & 0.57866 & 0.880 & 15 & 3 & 3 \\
3   & 0.63039 & 0.64135 & 0.992 &  7 & 3 & 3 \\
4   & 0.81130 & 0.81676 & 1.100 &  9 & 3 & - \\
5   & 0.87062 & 0.87602 & 1.085 & 17 & 3 & - \\
6   & 0.95085 & 0.95643 & 1.098 & 11 & 3 & - \\
\end{tabular}
\end{center}
\caption{Details of the LMC models. M$_\mathrm{1TP}$ and M$_\mathrm{TDUP}$ are the hydrogen exhausted core masses at the first thermal pulse and the first pulse for which TDUP occurs respectively. The maximum value of $\lambda$ obtained during the model run is $\lambda_\mathrm{max}$. The total number of thermal pulses evolved through is N$_\mathrm{tot}$, N$_\mathrm{TDUP}$ is the number of pulses prior to the occurence of TDUP and N$_\mathrm{c}$ is the number of pulses with TDUP required to form a carbon star. Note that stars of 4\ms\ or greater do not form carbon stars because of the occurence of hot-bottom burning.}
\label{tab:LMCmods}
\end{table}

For models of 2\ms\ or below the core masses at which the first episode of dredge-up occurs are lower than those in other models. For example, Karakas et al. (2002) find their 1\ms\ model without mass loss does not begin dredge-up until its core mass is 0.657\ms. This is 0.084\ms\ more massive than ours. This is important because we expect the lower-mass stars become the lower-luminosity carbon stars. We also find more efficient third dredge-up in these stars.

With the results of Karakas et al. (2002), Izzard et al. (2004) used a synthetic evolution code to fit the observed CSLF. They found that the minimum core mass for third dredge-up to occur would have to be lower by 0.07\ms\ and the minimum dredge-up efficiency would have to be 0.5 to get a good fit. Our results are in very good agreement with these predicitions. We find deeper and earlier dredge-up. Our models are also consistent with the requirement of a minimum temperature at the base of the convective envelope of $2.5\times10^{6}\,$K for third dredge-up to occur as used by Marigo, Girardi \& Bressan (1999) in their synthetic code.

\begin{figure}
\includegraphics[width=8cm]{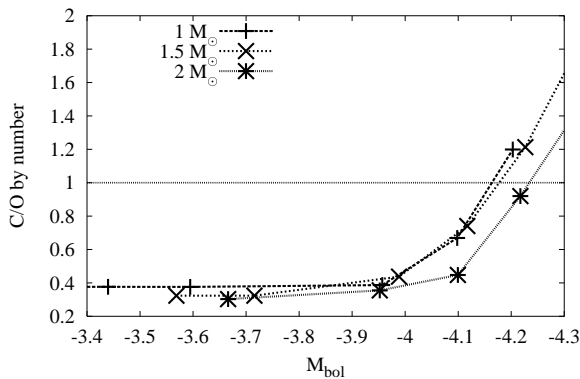}
\caption{The evolution of surface C/O with absolute bolometric magnitude, measured at the time of minimum luminosity in the interpulse, for models of 1, 1.5 and 2 M$_\odot$. The LMC carbon star luminosity function peaks at M$_\mathrm{bol}= - 4.9$.}
\label{fig:COMbol}
\end{figure}

Figure~\ref{fig:COMbol} shows the evolution of the surface C/O abundance by number with absolute bolometric magnitude. The data points are taken at the minimum luminosity occuring during the interpulse as this is what determines the lowest luminosity carbon star that forms. The results are encouraging. We find that our two lowest mass models are able to form carbon stars at an absolute bolometric magnitude of about $-4.2$. The LMC CSLF peaks at M$_\mathrm{bol} = -4.9$.

\subsection{SMC models}
As with the LMC models, we are able to form carbon stars in all our models below 3\ms. The details of the models are presented in Table~\ref{tab:SMCmods}. Again, carbon star formation follows rapidly once TDUP is established. For the same reasons as with the LMC model, $\lambda_\mathrm{max}$ is unlikely to be truely representative of the maximum efficiency reached.

\begin{table}
\begin{center}
\begin{tabular}{c|cccccc}
Mass & M$_\mathrm{1TP}$ & M$_\mathrm{TDUP}$ & $\lambda_\mathrm{max}$ & N$_\mathrm{tot}$ & N$_\mathrm{TDUP}$ & N$_\mathrm{c}$ \\ \hline
1   & 0.55929 & 0.58048 & 0.127 &  5 & 5 & 1 \\
1.5 & 0.57054 & 0.58849 & 0.774 & 12 & 4 & 2 \\
2   & 0.58133 & 0.60275 & 0.916 & 11 & 3 & 2 \\
3   & 0.69939 & 0.71421 & 1.048 &  9 & 3 & 2 \\
4   & 0.82842 & 0.83166 & 1.097 &  8 & 3 & - \\
5   & 0.90026 & 0.90292 & 1.068 & 14 & 3 & - \\
6   & 0.96795 & 0.97430 & 0.954 &  7 & 3 & - \\
\end{tabular}
\end{center}
\caption{Details of the SMC models. M$_\mathrm{1TP}$ and M$_\mathrm{TDUP}$ are the hydrogen exhausted core masses at the first thermal pulse and the first pulse for which TDUP occurs respectively. The maximum value of $\lambda$ obtained during the model run is $\lambda_\mathrm{max}$. The total number of thermal pulses evolved through is N$_\mathrm{tot}$, N$_\mathrm{TDUP}$ is the number of pulses prior to the occurence of TDUP and N$_\mathrm{c}$ is the number of pulses with TDUP required to form a carbon star. Note that stars of 4\ms\ or greater do not form carbon stars.}
\label{tab:SMCmods}
\end{table}

\begin{figure}
\includegraphics[width=8cm]{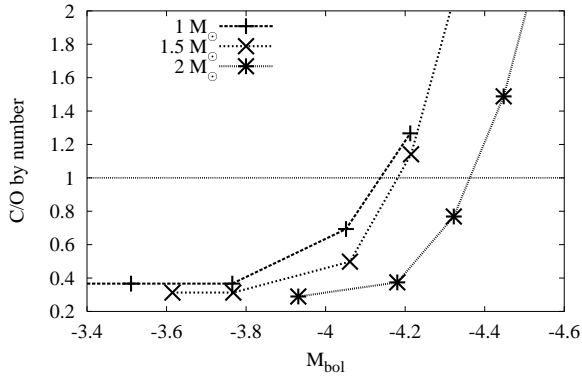}
\caption{The evolution of surface C/O with absolute bolometric magnitude, measured at the time of minimum luminosity in the interpulse, for models of 1, 1.5 and 2 M$_\odot$. The SMC carbon star luminosity function peaks at M$_\mathrm{bol}= - 4.5$.}
\label{fig:SMCCOMbol}
\end{figure}

The results of the SMC models runs are less encouraging from the point of view of reproducing the SMC carbon star luminosity function. Our low-mass models give minimum core masses for third dredge-up similar to those of Karakas et al. (2002). Our only advantage is the occurence of TDUP in the 1\ms\ model. Figure~\ref{fig:SMCCOMbol} shows the evolution of the C/O abundance with absolute bolometric magnitude. The lowest mass models become carbon stars with M$_\mathrm{bol}=-4.2$. This magnitude is only just below the peak of the SMC CSLF.

\section{Population Synthesis}
In order to see if we can reproduce the LMC and SMC CSLF we need to create a population of stars based on our detailed models. We utilise the synthetic TP-AGB evolution code of Izzard et al. (2004). The luminosity core-mass relation therein is found to fit our data well and so does not need to be refitted. As in Izzard et al. (2004), the luminosity dip after each thermal pulse (see Iben \& Renzini 1983) is modelled by a factor of the form
\be
f_\mathrm{L} = 1 - 0.5 \times \mathrm{min}\left[1,\exp(-3 {\tau\over \tau_\mathrm{ip}})\right],
\ee
\begin{figure}
\includegraphics[width=8cm]{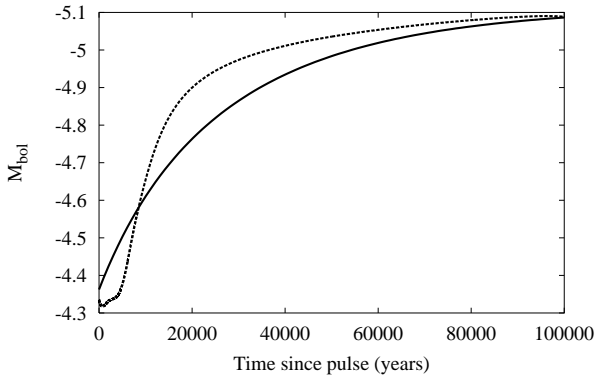}
\caption{Plot to of the exponential fit (solid line) to the post-flash luminosity dip of the 4th pulse of the 2\ms\ Z$ = 0.008$ model (dashed line). The fit leads to an underestimate of the bolometric magnitude of about 0.1 mag.}
\label{fig:pulsefit}
\end{figure}
where $\tau$ is the time from the beginning of the current pulse and $\tau_\mathrm{ip}$ is the interpulse period. In this form the luminosity dip found in the detailed models can be reproduced (see Figure~\ref{fig:pulsefit}). It is necessary to include this dip in order to reproduce the low-luminosity side of the CSLFs. The core masses at the first thermal pulse, $M_\mathrm{1TP}$, minimum core mass for third dredge-up to occur, $M_\mathrm{c,min}$, and $\lambda$ values used by the synthetic code are fitted to the above detailed models. Additional models of 1.25, 1.75 and 2.25\ms were made for the LMC in order to better model the behaviour of the core mass. The fits are
\begin{eqnarray}
 M_\mathrm{1TP} = 0.9557 + 0.31741\exp\left[-{(M_*-1.164)^2\over1.1192}\right] \nonumber \\ -{0.82352\over1+0.31831^{2.7546-M_*}}
\end{eqnarray}
for the LMC and
\be
 M_\mathrm{1TP} = 0.56633 + {0.067603M_*\over1+0.096788^{M_*-2.6354}}
\ee
for the SMC, where $M_*$ is the mass of the star at the beginning of the TP-AGB. Third dredge-up begins in the models when the core mass reaches a minimum value, $M_\mathrm{c,min}$, defined by
\begin{eqnarray}
 M_\mathrm{c,min} = 0.9659 + 0.26309\exp\left[-{(M_*-1.0724)^2\over1.2629}\right] \nonumber \\ -{0.73763\over 1 + 0.3344^{2.86280-M_*}}
\end{eqnarray}
for the LMC and
\be
 M_\mathrm{c,min} = 0.58042 + {0.065376M_*\over1+0.098049^{M_*-2.6691}}
\ee
for the SMC.

The behaviour of $\lambda$ once TDUP has commenced is modelled by
\be
\lambda = a[1-(1-0.42n)\mathrm{e}^{-\frac{n}{b}}],
\ee
\begin{eqnarray}
a = \mathrm{max}(0,2.0369 + 15Z - 2.8775M_* + 1.6206M_*^2 \nonumber \\ - 0.335270M_*^3 + 0.023211M_*^4),
\end{eqnarray}
\be
b = 6.835 - {4.44870\over 1+0.002^{M_*-2.5}},
\ee
where $Z$ is the metallicity and $n$ is the pulse number counting the first pulse where dredge-up occured as 1. The last factor in Equation (4) allows $\lambda$ to rise to a maximum over several pulses and then begin to fall off as is seen in the detailed models. The low number of thermal pulses calculated in the 1\ms\ models makes a reliable estimate for $\lambda$ difficult and so for stars of between 1 and 1.5\ms\ we use the values as computed for the 1.5\ms\ case. This makes dredge-up slightly more efficient but as carbon stars form rapidly at such low masses this does not significantly affect the results.

A population of 10,000 stars was evolved for 16 Gyr. A Kroupa, Tout \& Gilmore (1993) initial mass function and a constant star formation rate were assumed. Mass loss was included as the Vassiliadis \& Wood (1993) type prescription used by Karakas et al. (2002). The superwind phase is turned on when the Mira pulsation period of the star reaches $500\,$d. We have fitted the synthetic code to results from detailed models without mass loss. Because the Vassiliadis \& Wood formalism causes significant mass loss only in the later pulses and our models form carbon stars rapidly, the impact of this approximation on our results should be limited.

\begin{figure}
\includegraphics[width=7cm,angle=270]{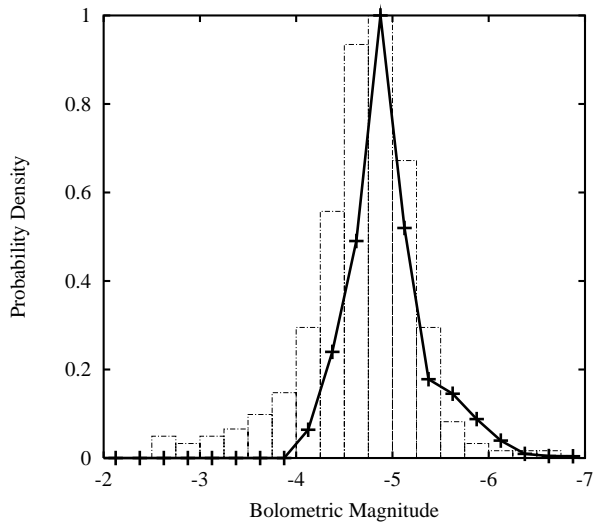}
\caption{The theoretical fit (solid line) to the LMC CSLF. The histogram is observational data taken from Groenewegen (2002). The CSLF is reasonably well reproduced by the theoretical models.}
\label{fig:LMCfit}
\end{figure}

\begin{figure}
\includegraphics[width=7cm,angle=270]{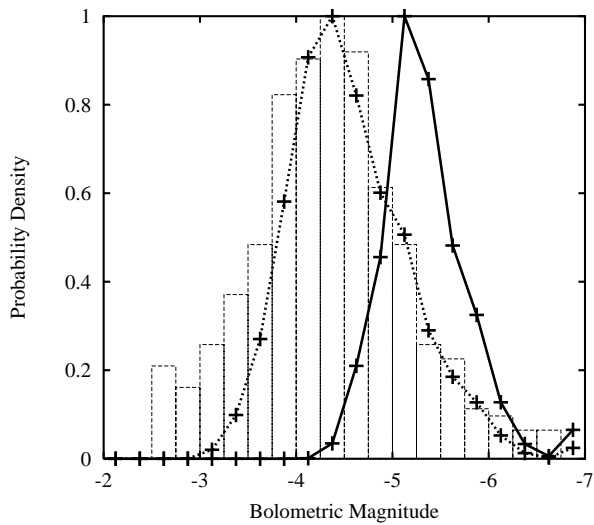}
\caption{Theoretical fits to the SMC CSLF. The histogram is observational data taken from Groenewegen (2002). The unadjusted theoretical model (solid line) reproduces the shape of the luminosity function well, but it is too bright by almost 1 magnitude. If the minimum core mass required for TDUP is reduced by 0.06\ms\ (dashed line) then the CSLF is well reproduced.}
\label{fig:SMCfit}
\end{figure}

The results of the population synthesis runs are shown in Figures~\ref{fig:LMCfit} and~\ref{fig:SMCfit}. The models are normalised such that the peak of a model matches the peak of the corresponding observations. The model fits the observations of the LMC CSLF very well except for the model a slight underabundance of carbon stars between M$_\mathrm{bol}=-4$ and M$_\mathrm{bol}=-4.75$. Note that we do not fit the very low-luminosity carbon stars. These are likely to be extrinsic, rather than intrinsic, carbon stars (see Izzard \& Tout 2004).

The fit to the SMC CSLF is somewhat disappointing. The peak in the model luminosity function is almost 1 magnitude too bright. If the minimum core mass for dredge-up is reduced by 0.06 \ms\ as in Izzard \& Tout (2004) then we are able to reproduce the SMC CSLF (see Figure~\ref{fig:SMCfit}). Again, the very low luminosity tail is expected to be due to extrinsic carbon stars.

It should be noted that our models consist only of a single metallicity and do not account for the finite size of the clouds. Both these effects would broaden the luminosity function slightly. The work of Harries, Hilditch \& Howarth (2003) gives examples of systems with distance moduli that can be as much as 0.2 magnitudes from the quoted distance to the SMC. This is unlikely to have a great effect on the luminosity function, given the binning is of this order but it may improve the LMC model. The effect of metallicity variations is much more difficult to quantify, though we note that varying the metallicity by a factor of 2 (i.e. going from Z$=0.008$ for the LMC to Z$=0.004$ for the SMC) does produce noticeably different results.

\section{Conclusion}
We have made evolution calculations, with a fully implicit and simultaneous code, of the thermally pulsing asymptotic giant branch evolution of low- and intermediate-mass stars at LMC and SMC metallicities. For the LMC we find more efficient dredge-up at lower core masses than others. The SMC models are comparable to those of Karakas et al. (2002). Using a population synthesis code calibrated to the detailed models we find we are able to reproduce the carbon star luminosity function of the LMC. The SMC CSLF can still only be reproduced if the minimum core mass required for TDUP is reduced by 0.06\ms.

The fact that these calculations can reproduce, for the first time, the
LMC CSLF is of great interest.  It is not clear why the present calculations succeed while others have failed, but the simultaneous solution of structure and mixing may be important. We are currently exploring the reasons for the difference between our results and those of others.

\section{Acknowledgements}
We thank the anonymous referee for their helpful comments. RJS thanks PPARC for a studentship. RGI thanks CIQuA for computing and monetary support. CAT thanks Churchill College for a Fellowship.

\label{lastpage}

\end{document}